\documentclass[aps,prd,nobibnotes,twocolumn,superscriptaddress,bibliography]{revtex4-2}

\usepackage{amsfonts}
\usepackage{mathrsfs}
\usepackage{amsmath}
\usepackage{color}
\usepackage{graphicx}
\usepackage{bm}
\usepackage{amssymb}
\usepackage{xspace}
\usepackage{epstopdf}
\usepackage{dcolumn}
\usepackage{longtable}
\usepackage{multirow}
\usepackage{float}
\usepackage{comment}
\usepackage{braket}
\usepackage{soul}
\usepackage{multirow}

\usepackage{listings}
\usepackage{xcolor}
\usepackage{mathematica}
\definecolor{darkred}{rgb}{0.5 0 0}
\definecolor{darkgreen}{rgb}{0.5 .5 0}
\definecolor{darkblue}{rgb}{0 0 .5}

\providecommand{\tabularnewline}{\\}

\usepackage[colorlinks=true, letterpaper=true, pdfstartview=FitV, linkcolor=blue, citecolor=blue, urlcolor=blue]{hyperref}

\makeatother
\begin{document}
	\title{A phonon irreducible representations calculator}

\author{Zeying Zhang}
\affiliation{College of Mathematics and Physics, Beijing University of Chemical
	Technology, Beijing 100029, China}
\author{Zhi-Ming Yu}
\affiliation{Centre for Quantum Physics, Key Laboratory of Advanced Optoelectronic
	Quantum Architecture and Measurement (MOE), School of Physics, Beijing
	Institute of Technology, Beijing, 100081, China}
\affiliation{Beijing Key Lab of Nanophotonics \& Ultrafine Optoelectronic Systems,
	School of Physics, Beijing Institute of Technology, Beijing, 100081,
	China}
\author{Gui-Bin Liu}
\email{gbliu@bit.edu.cn}
\affiliation{Centre for Quantum Physics, Key Laboratory of Advanced Optoelectronic
	Quantum Architecture and Measurement (MOE), School of Physics, Beijing
	Institute of Technology, Beijing, 100081, China}
\affiliation{Beijing Key Lab of Nanophotonics \& Ultrafine Optoelectronic Systems,
	School of Physics, Beijing Institute of Technology, Beijing, 100081,
	China}
\author{Yugui Yao}
\email{ygyao@bit.edu.cn}

\affiliation{Centre for Quantum Physics, Key Laboratory of Advanced Optoelectronic
	Quantum Architecture and Measurement (MOE), School of Physics, Beijing
	Institute of Technology, Beijing, 100081, China}
\affiliation{Beijing Key Lab of Nanophotonics \& Ultrafine Optoelectronic Systems,
	School of Physics, Beijing Institute of Technology, Beijing, 100081,
	China}
\begin{abstract}
The irreducible representation of band structure is important for physical properties. Based on phonopy and recently developed SpaceGroupIrep package, we developed a package PhononIrep, which can get the band irreducible representation for arbitrary $\boldsymbol{k}$ point at first-principles level. As an application, we for the first time predict the  cubic crossing Dirac point can exist in conventional crystal phonon systems  X$_3$Y (X=Ti, Nb, Ta, Y=Au, Sb).  We hope this package will facilitate phonon research in the future.
\end{abstract}
\maketitle
	
\text{\it Introduction.}
Group representation theory is a powerful tool to investigate  novel properties in theoretical physics. In solid state
physics, representation theory can not only greatly simplify the first-principles calculations \cite{kresse_efficient_1996}, but also can help  us to investigate the connectivity of energy bands in crystals \cite{michel_connectivity_1999}, the lattice vibrations \cite{lax_symmetry_2001}, the selection rules \cite{bradley_mathematical_2009}, the effective Hamiltonians \cite{kane_band_1957, egorov_consistent_1968, tang_exhaustive_2021, jiang_kp_2021,zhang_magnetictb_2021}.  Very recently, researchers use group representation theory to search topological insulators and emergent particles \cite{bradlyn_beyond_2016,bradlyn_topological_2017,po_symmetry-based_2017,yu_encyclopedia_2021,tang_comprehensive_2019,zhang_catalogue_2019,vergniory_complete_2019,zhang_research_2020,wang_coexistence_2021,zhou_hybrid-type_2021,xie_sixfold_2021,liu_encyclopedia_2021,zhang_encyclopedia_2021}.

At present,  there are four packages IrRep, irvsp, qeirreps and SpaceGroupIrep \cite{iraola_irrep_2020,liu_spacegroupirep_2021,gao_irvsp_2021,matsugatani_qeirreps_2021} focus on space group irreducible representations. The data in first three packages are compatible with BCS website \cite{aroyo_bilbao_2006}, and in SpaceGroupIrep is compatible with Bradley and Cracknell's (BC) book \cite{bradley_mathematical_2009}. However, those packages are mainly used for electronic systems not for the phonon systems except irvsp can determine the phonon irreducible representations in tight-binding level \cite{gao_irvsp_2021}. In addition, the phonopy  package can only give the  irreducible representations of $\Gamma$ point but not for general $\boldsymbol{k}$ point \cite{togo_first_2015, togo_spglib_2018}.

In this work, We make an expansion of SpaceGroupIrep, namely, PhononIrep, which can get the band irreducible representation for arbitrary $\boldsymbol{k}$ point in first-principles level.
The only necessary inputs of PhononIrep is the force constant from Phonopy and the structure of unit cell . As an example, we for the first time proposed the cubic crossing Dirac point (CCDP) 
in phonon system.  PhononIrep 
will stimulate further studies on the fascinating properties for phonon system.

\text{\it Method.}
In current version of phonopy, the  irreducible representations can be obtained for only $\Gamma$ point, but for general $\boldsymbol{k}$  points  phonopy can only calculate the  character $\chi^{l}_{\boldsymbol{k}}(Q)$ of $\boldsymbol{k}$.
Then $\chi^{l}_{\boldsymbol{k}}(Q)$ can be completely reduced by orthogonality relationships  
\begin{equation}
	\chi^{l}_{\boldsymbol{k}}(Q)=\sum_{i}a_i \chi^{i}_{\boldsymbol{k}}(Q)
	\label{eq:o}
\end{equation}
where $Q=\{R|\boldsymbol{t}\}$ with rotation part $R$ and translation part $\boldsymbol{t}$ is an element of the little group of $\boldsymbol{k}$, $l$ is the band index.
\begin{figure}[htbp]
	\includegraphics[width=\columnwidth]{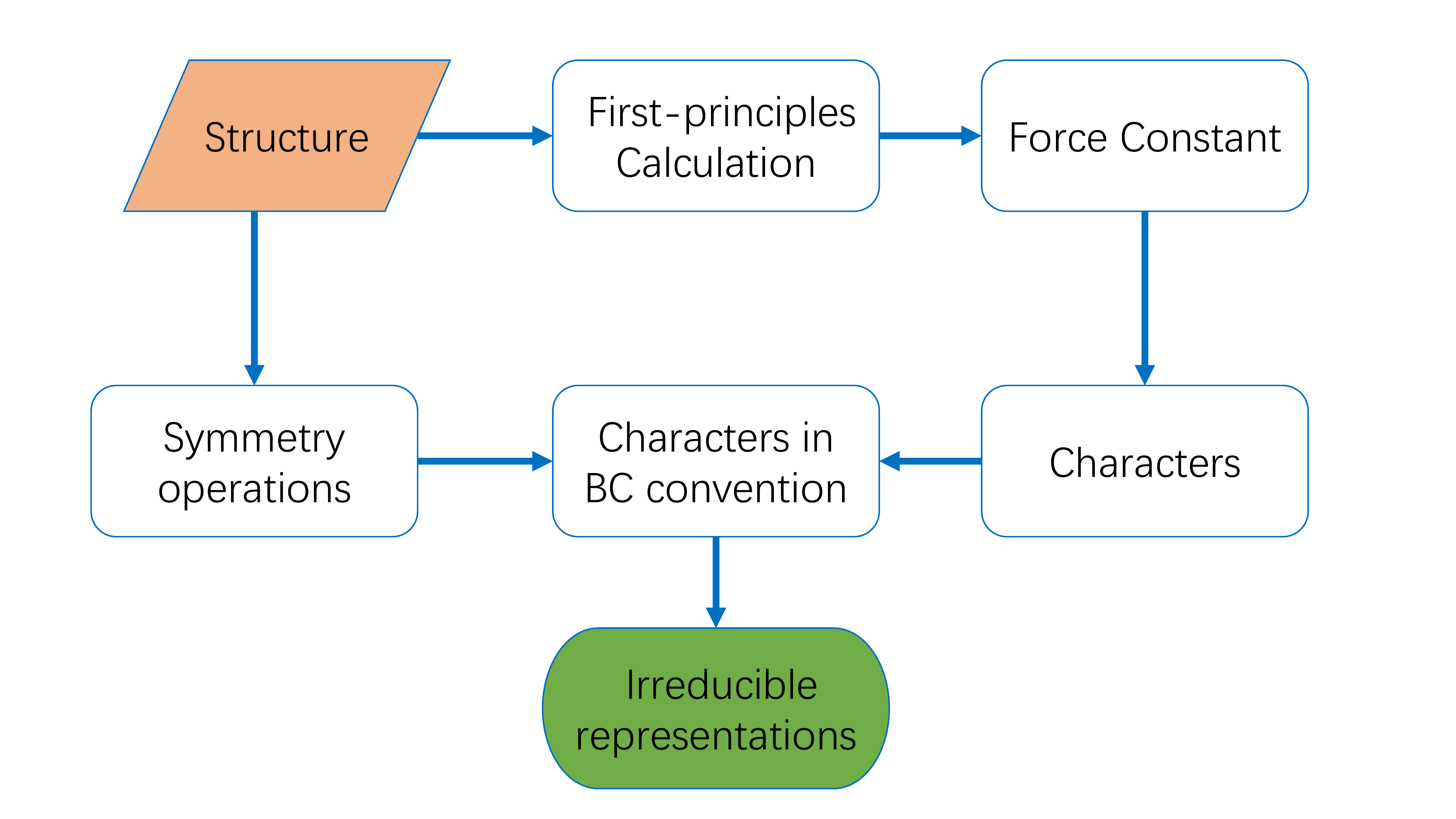}
	\caption{Flowchart of calculating the irreducible representations of phonon.
	\label{fig1}}
\end{figure}

The procedures for calculating the irreducible representations of phonon are as follow (Fig.(\ref{fig1})): Start form a unit cell structure. First, use finite displacement method or density functional perturbation theory to get the second-order force
constants. Second, use phonopy to calculate the characters of $Q$. Third, convert the 
character to BC convention. This step is done by convert the input cell and irreducible representation's labels into BC conventions \cite{liu_spacegroupirep_2021}.
Finally, calculate the irreducible representations by comparing the obtained characters and the character table of the little group of $\boldsymbol{k}$.

Comparing to irvsp, PhononIrep have the following advantages:
\begin{itemize}
	\item Calculate phonon  irreducible representations from first-principles results directly, there is no need to convert the force constants into tight-binding Hamiltonian.
	\item There is no need to standardise the unit cell structure manually by phonopy or spglib before the first-principles calculation.
	\item The label of irreducible representations is given for both Mulliken \cite{mulliken_electronic_1933} and $\Gamma$ label \cite{bouckaert_theory_1936,koster_properties_1963,bradley_mathematical_2009}.
\end{itemize}

\begin{figure}[tbp!]
	\includegraphics[width=\columnwidth]{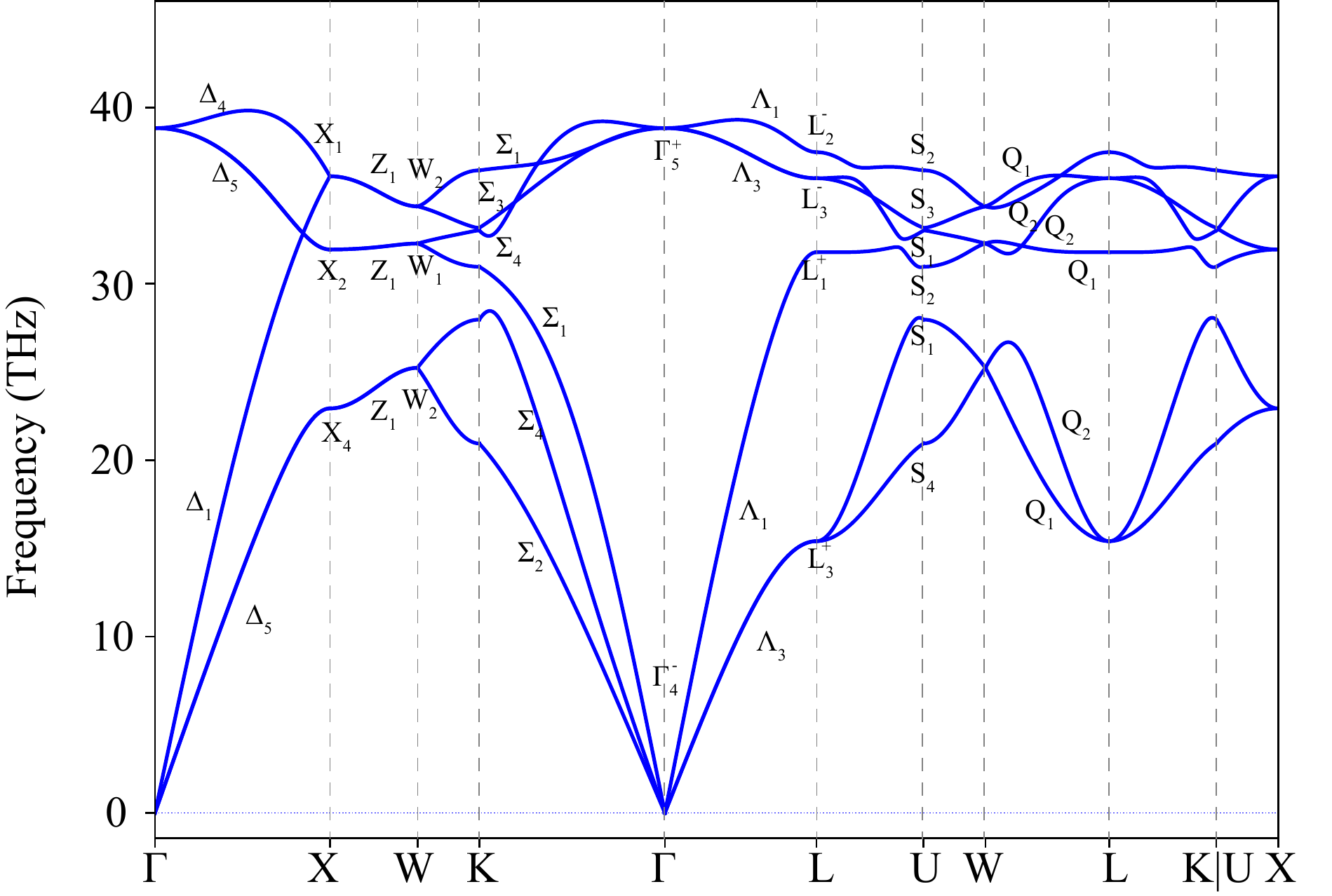}
	\caption{Phonon spectrum of diamond. The little representations are labelled by $\Gamma$ notation.}
	\label{fig:diamond}
\end{figure}

The inputs of PhonoIrep are clearly and easy to set. In principle, only the structure and force constants information are
sufficient to  get the irreducible representations.
One can run the following short script to get the irreducible representations of  phonon system:
\begin{lstlisting}[backgroundcolor={\color{yellow!5!white}},mathescape=true]
calcPhononIrep["supercell" -> size of supercell,
"unitcell" -> path of unit cell file,
"force" ->  path of force constants file,
"kset"-> list of $\boldsymbol{k}$ points, 
"symprec" -> $\epsilon_1$,
"degeneracytolerance" -> $\epsilon_2$,
"showRep"->True or False]
\end{lstlisting}
\lstinline!calcPhononIrep! is the function to get the irreducible representations which have six options. 
\lstinline!"supercell"! is a list including three integers represent the size of the supercell in phonon calculation.
\lstinline!"unitcell"! is the full path of unit cell structure file. \lstinline!"force"! is the full path of unit force constants file.  \lstinline!"kset"! is a $n\times3$  array that tell PhononIrep
which $\boldsymbol{k}$ points will be calculated. \lstinline!"showRep"! tell PhononIrep whether to show the 
results in a \lstinline!Grid! form. \lstinline!"symprec"! and \lstinline!"degeneracytolerance"! are tolerance of determining the symmetry and the degeneracy in phonopy, the default values of \lstinline!"symprec"! and \lstinline!"degeneracytolerance"! are 0.0001, which are fine for most of calculations (see Appendix~\ref{app1} for installation and running of PhononIrep).

As an example, we use PhononIrep to calculate  irreducible representations of diamond (Fig.(\ref{fig:diamond})). The  irreducible representations are consistent with Ref.~[\onlinecite{lax_symmetry_2001}]. Therefore, PhononIrep can avoid tedious derivation and get the result directly.

\begin{figure}[tbp]
	\includegraphics[width=\columnwidth]{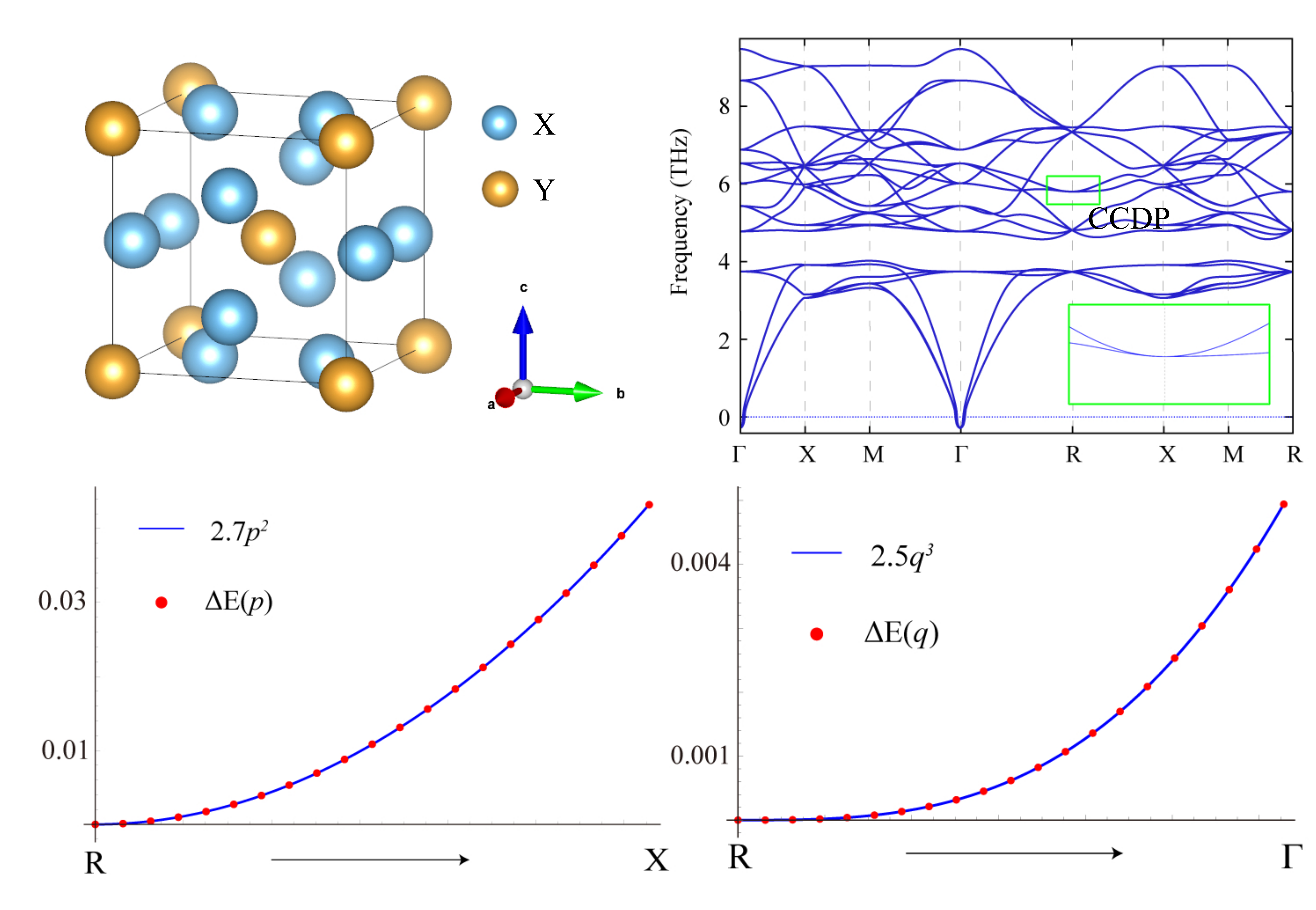}
	\caption{(a) Crystal structures of  X$_3$Y. (b) The phonon dispersion of Ti$_3$Au along high symmetry path,
		the CCDP
		marked within the green rectangles are zoomed in around the zero frequency. (c-d) The frequency difference of CCDP along different path, along path $R-X$ the dispersion are quadratic, along path $R-\Gamma$ the  dispersion are cubic.
	}
	\label{fig:x3y}
\end{figure}

\text{\it An application---CCDP.} The CCDP  with cubic leading order dispersion on the four body diagonals of the cube Brillouin zone  is merged by two C-4 Weyl points.
According to the encyclopedia of emergent particles \cite{yu_encyclopedia_2021}, CCDP can only appear in cubic structure with space group 218, 220, 222, 223, 230 in spinless system. The real materials which host CCDP has not  been reported before.
Here, we take Ti$_3$Au as an example to confirm the CCDP can exist in phonon system. 
Ti$_3$Au  host simple cubic structure. Ti atom locate at $6c$ Wyckoff position and Au locate at $2a$ Wyckoff position. The crystal structure of Ti$_3$Au is shown in Fig.~\ref{fig:x3y}(a). After the force constants is calculated, we can immediately get the irreducible representations of $R(\frac{1}{2},\frac{1}{2},\frac{1}{2})$ by PhononIrep. As shown in table~\ref{tab:ir}, three exist three six-fold, one  and one four-fold degeneracy points. The four-fold point is a CCDP which locate at 5.80 THz formed by $R_2\oplus R_3$.  One can easily verify the degeneracies by checking the phonon dispersions [see Fig.~\ref{fig:x3y}(b)]. 
\begin{table}
	\caption{Irreducible representations of $R(\frac{1}{2}\frac{1}{2}\frac{1}{2})$ point for Ti$_3$Au. The first column show the band number of $n-$th band.  The unit of frequency is THz, ``dim'' represent the dimensions of the irreducible representations. The label of irreducible representations are given  in Mulliken and $\Gamma$ label.}
	\begin{tabular}{@{\extracolsep{\fill}}*{1}{cccccc}@{}}
		\hline
		\hline
		\multicolumn{2}{c}{\multirow{1}{*}{band}}  &
		\multicolumn{1}{c}{\multirow{2}{*}{frequency}} & 
		\multicolumn{1}{c}{\multirow{2}{*}{dim}} &
		\multicolumn{2}{c}{\multirow{1}{*}{irreducible representations}}     \tabularnewline
		\cline{1-2}\cline{5-6}
		start&end&&&Mulliken&$\Gamma$ label \tabularnewline
		\hline
		1&6  & 3.74 & 6 &$H$ & $R_4$  \tabularnewline
		7&12 & 4.81	&6&$H$&$R_4$  \tabularnewline
		13&16& 5.80 &4&\, $^1F\oplus^2F$ &$R_2\oplus R_3$ \tabularnewline
		17&22& 7.34 &6&$H$ & $R_4$  \tabularnewline
		23&24& 7.46 &2&$E$&$R_1$ \tabularnewline
		\hline
	\end{tabular}
	\label{tab:ir}
\end{table}

Within the calculated irreducible representations, we can do a more detailed analysis of this material. 
A set of generators and representation matrices  of the symmetry operators for $R$ point are given by 
\begin{equation}
\begin{split}
C_{2z}=&C_{2x}=\Gamma _{0,0}, I= \Gamma _{0,3}\\
C_{31}^+=& -\frac{\Gamma _{0,0}}{2}-\frac{1}{2} i \sqrt{3} \Gamma _{3,3}\\
C_{2z}=&\frac{1}{2} \sqrt{3} \Gamma _{3,2}-\frac{\Gamma _{0,1}}{2}\\
{\cal T}=& \Gamma _{1,0}\\
\end{split}
\end{equation}
where $\Gamma_{i,j}=\sigma_i\otimes\sigma_j$ and $\sigma_i (i=0,1,2,3)$ are identity matrix and three Pauli matrices.

We then can construct  the $\boldsymbol{k}\cdot\boldsymbol{p}$ model of $R$ point that satisfies the symmetry constrains 
\begin{equation}
P(Q)H(k)P^{-1}(Q)=H(Q\boldsymbol{k})
\end{equation}
where $P(Q)$ is the representation matrix of the little group for $R$. Consider the Hamiltonian upto third order. The obtained model can be written as

\begin{equation}
	\begin{split}
		&H=\begin{pmatrix}
			  \lambda^* \alpha  k_x^2+\alpha k_y^2+\lambda \alpha k_z^2 & \beta k_x k_y k_z \\
			-\beta k_x k_y k_z & \lambda^* \alpha  k_x^2+\lambda \alpha  k_y^2+\alpha k_z^2 \\
		\end{pmatrix}\\
	&\\
	&H^{223,R}_\text{CCDP}=\varepsilon+ck^2+\begin{pmatrix}
		0 & H\\
		H^\dagger &0
	\end{pmatrix}
	\end{split}
		\label{eq:ham}
	\end{equation}
Here,  $\lambda=e^{\frac{2 i \pi }{3}}$, $k^2=k_x^2+k_y^2+k_z^2$, $\varepsilon, c$ are real parameters and $\alpha, \beta$ are complex parameters. Notice in four body diagonals of the cube Brillouin zone, the leading order dispersion in the
band splitting is cubic, i.e. for $q^2=k_x^2=k_y^2=k_z^2$, the spectrum of this Hamiltonian (\ref{eq:ham}) is
\begin{equation}
E(q)=\varepsilon+3cq^2\pm|\beta| q^3 
\end{equation}
where the $-(+)$ represent the  lower (higher) two bands of CCDP.  
However, in six face diagonals the dispersion are quadratic, e.g. along R$-$X high symmetry, $p=k_x=k_z,k_y=0$, the 
spectrum of this Hamiltonian (\ref{eq:ham}) is
\begin{equation}
	E(q)=\varepsilon+2cp^2\pm|\alpha| p^2 
\end{equation}
which is different from the conventional cubic Dirac point composed of two C-3 Weyl fermions \cite{yang_classification_2014}. Actually the CCDP composed of two C-4 Weyl fermions, when the time reversal symmetry or
inversion is broken, the CCDP will split into two  C-4 Weyl fermions with opposite topological charge  \cite{cui_charge-4_2021}. Such difference will play a
dominant role in the physical properties of the system \cite{yu_quadratic_2019}.

For the phonon calculations of diamond and Ti$_3$Au, the second-order force constants is based on the density functional perturbation theory within the phonony and VASP package \cite{kresse_efficient_1996, togo_first_2015}. A $2\times2\times2$ supercell of the conventional unit cell with $3\times3\times3$ $k-$mesh and 300 eV for energy cutoff are applied.
It should be emphasized that CCDP always exist in proper frequency for materials with space group 218, 220, 222, 223, 230.  Therefore, we partially list a series of materials  X$_3$Y (X=Ti, Nb, Ta, Y=Au, Sb) in Materials project \cite{jain_commentary_2013} (as shown in Appendix~\ref{app2}) which have same symmetry with Ti$_3$Au. Such materials must
have CCDP in $R$ point of Brillouin zone.

In conclusion, we have developed a package PhononIrep which can get the irreducible representations of phonon system. As an application, it confirms that the CCDP can exist in realistic phonon systems. Our work provide a useful tool to analyse the phonon system. An exciting direction for future is to use PhononIrep to search for topological materials.

\text{\it Acknowledgments.}
This work is supported by
the NSF of China (Grants Nos.~12004028, 12004035, 11734003, 12061131002),
the China Postdoctoral Science Foundation (Grant No.~2020M670106),
the Strategic Priority Research Program of Chinese
Academy of Sciences (Grant No.~XDB30000000), the National Key
R\&D Program of China (Grant No.~2020YFA0308800),
, and the Beijing Natural Science Foundation (Grant No.~Z190006).


\bibliography{PhononIrep}

\onecolumngrid

\appendix

\section{Installation and running of PhononIrep}
\label{app1}
Before install PhononIrep,  users should install SpacegroupIrep and Phonopy.
Then, the steps of installing PhononIrep is exactly same as installing MagneticTB \cite{zhang_magnetictb_2021}.
Unzip the "PhononIrep-main.zip" file and copy
the PhononIrep directory to any of directory in \lstinline!$Path!. e.g.
copy the PhononIrep directory to \lstinline!FileNameJoin[{$UserBaseDirectory, "Applications"}]!.
Then one can use the package after running \lstinline!Needs["PhononIrep`"]!.
The version of Mathematica should $\ge$ 11.2. 
The source code of PhononIrep are available on \url{https://github.com/zhangzeyingvv/PhononIrep}.

One can run the following script to get the irreducible representations of $R$ point for Ti$_3$ Au
\begin{lstlisting}[backgroundcolor={\color{yellow!5!white}},mathescape=true,numbers=left]
SetEnvironment["PATH" -> Environment["PATH"] <> ";" <> "D:\\Anaconda3\\Library\\bin"];
RegisterExternalEvaluator["Python", "D:\\Anaconda3\\python.exe"];
FindExternalEvaluators["Python"];
Needs["PhononIrep`"];
calcPhononIrep["supercell" -> {2, 2, 2},
"unitcell" -> NotebookDirectory[] <> "POSCAR-unitcell",
"force" ->  NotebookDirectory[] <> "FORCE_CONSTANTS",
"kset" -> {{0.5, 0.5, 0.5}}, "showRep" -> True]
\end{lstlisting}

Line 1-3 is to set the Python environment in Mathematica. Users have to specify the Python installation path in \lstinline!RegisterExternalEvaluator! and \lstinline!SetEnvironment!, see \url{https://support.wolfram.com/52852} for detail. 
Line 4 is to load PhononIrep package.
See main text for a detailed description of line 5-8.
The output of above script is:

\begin{figure}[H]
\centering
\includegraphics[width=0.5\columnwidth]{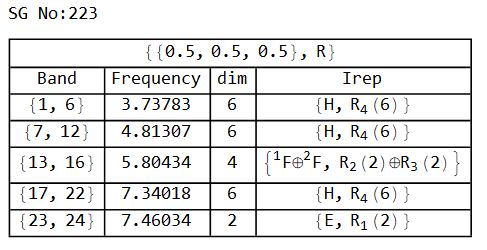}
\end{figure}
To directly compare the irreducible representations between the output of PhononIrep and the results in Ref.~[\onlinecite{yu_encyclopedia_2021}], one can use \lstinline!checkLGIrepLabel! in SpaceGroupIrep to get the abstract groups' irreducible representations label:
\begin{lstlisting}[backgroundcolor={\color{yellow!5!white}},mathescape=true]
Grid[checkLGIrepLabel[223, "R"], Frame -> All]
\end{lstlisting}
the output of above command is:
\begin{figure}[H]
	\centering
	\includegraphics[width=0.3\columnwidth]{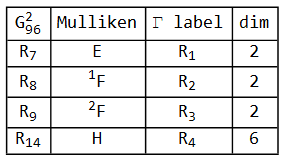}
\end{figure}

\section{List of X$_3$Y}
\label{app2}
\begin{table}[h]
	\caption{List of mp-ID, lattice constant and ICSD for X$_3$Y (X=Ti, Nb, Ta, Y=Au, Sb), the structure data are available on \url{www.materialsproject.org}.}
	\begin{tabular}{@{\extracolsep{\fill}}*{1}{c|c|c|c}@{}}
		\hline
		\hline
		Material & mp-ID	&lattice constant (\AA) & ICSD(s) \tabularnewline
		\hline
		Ti$_3$Au &mp-1786	&5.113&612419 612405 58605 612417 612420 612418    \tabularnewline
		Nb$_3$Au &mp-2752	&5.256&612199 58557 612192 612186 612203 612188 612198 612185 \tabularnewline
		Ta$_3$Au &mp-569249	&5.246&58599\tabularnewline
		V$_3$Au  &mp-839	&4.881&612456 58612 612451 612459  \tabularnewline
		Ti$_3$Sb &mp-1412	&5.217&651683 96137 651684 106035 651685 43356 657034 \tabularnewline
		Nb$_3$Sb &mp-2053	&5.312&645347 76572 645349 190178 645357 645352 \tabularnewline
		Ta$_3$Sb &mp-541	&5.303&52310 651601\tabularnewline
		V$_3$Sb  &mp-1555	&4.935&651731 52330 106037 651717  \tabularnewline
		\hline
	\end{tabular}
	\label{tab:mats}
\end{table}

\end{document}